\documentclass[aip, amsmath, amssymb, reprint]{revtex4-1}
\usepackage{graphicx}
\usepackage{dcolumn}
\usepackage{bm}
\usepackage[utf8]{inputenc}
\usepackage[T1]{fontenc}
\usepackage{mathptmx}
\usepackage{etoolbox}
\usepackage{textgreek}      
\usepackage{xr}
\makeatletter
\def\@email#1#2{
 \endgroup
 \patchcmd{\titleblock@produce}
  {\frontmatter@RRAPformat}
  {\frontmatter@RRAPformat{\produce@RRAP{*#1\href{mailto:#2}{#2}}}\frontmatter@RRAPformat}
  {}{}
}
\makeatother
\begin{document}
\preprint{AIP/123-QED}

\title[]{A persistent-current-biased and current-actuated switch for superconducting circuits}

\author{Ziyi Zhao} \thanks{Author to whom correspondence should be addressed: zz544@yale.edu}
    \affiliation{JILA, National Institute of Standards and Technology and the University of Colorado, Boulder, CO, USA.}
    \affiliation{Department of Physics, University of Colorado, Boulder, CO, USA.}
\author{Eva Gurra}
    \affiliation{JILA, National Institute of Standards and Technology and the University of Colorado, Boulder, CO, USA.}
    \affiliation{Department of Physics, University of Colorado, Boulder, CO, USA.}
\author{Michael R. Vissers}
    \affiliation{National Institute of Standards and Technology, Boulder, CO, USA.}
\author{K. W. Lehnert}
    \affiliation{Department of Physics, Yale University, New Haven, CT, USA.}

\date{\today}

\begin{abstract}
Broadband and low-loss superconducting switches can facilitate large-scale quantum information processors and cryogenic detectors by dynamically reconfiguring the connectivity of their circuits. The time dependent connectivity is enabled by the nonlinearity of lossless Josephson junctions, which are often wired into superconducting loops to be controlled by magnetic flux. However, this approach needs a power-consuming constant flux bias and dynamic flux actuation, both of which are hard to isolate from other switches or flux sensitive elements, limiting their integration density.
Here, we design and characterize a microwave switch that implements a persistent current bias and direct current actuation to reduce static power consumption, actuation energy and potential crosstalk to other devices. 
We show that persistent current associated with tens of flux quanta is stable and long-lived, reducing the need for on-the-fly tuning. 
We  further demonstrate that our switch has desirable performance for superconducting-circuit-based quantum information processing, including an off mode with more than $ 20 \ \text{dB}$ isolation comparable to commercial ferrite isolators, power handling larger than $100\ \text{pW}$ sufficient for resonator readout tones and amplifier pumps, and modulation bandwidth broader than $ 600\ \text{MHz}$ useful for multiplexing schemes.
\end{abstract}
\maketitle

Cryogenic microwave measurements require signal routing and multiplexing capabilities to facilitate large-scale and modular designs. Both requirements can be addressed in part by switches that rapidly modulate. For example, in modular quantum networks, switches are connected to high quality-factor modules where quantum states are stored. By actuating particular switches, quantum information can then be routed to desired modules, enabling quantum gates, state readout or other operations \cite{Error-Detected_Luke_2021, High-On-Off-Ratio_Chapman_2023, Reaching_Reager_2013, Three-Dimensional_Romanenko_2020}. In quantum random access memory and qubit control schemes, switches are used as multiplexers to route signals to the desired modes \cite{Random_Naik_2017, Hardware-Efficient_Hann_2019, Cryogenic_Huang_2022, Realizing_Zhou_2023}. Similarly, in large-format cryogenic detectors, switches can dynamically re-allocate the shared readout resources \cite{Advanced_Irwin_2012, Kilo-pixel_Baselmans_2017, Superconducting_Oripov_2023}. Lastly, switches facilitate in-situ calibration of microwave power \cite{Two-port_Ranzani_2013}.

In these applications, a versatile switch should be broadband, low-loss and easy to actuate. Furthermore, in large scale applications, they should be compatible with high density integration. Devices designed for this purpose have been proposed or implemented in various ways \cite{Operation_Borodulin_2019, Demonstration_Wagner_2019, Control_Lecocq_2021, Multiplex_Acharya_2023, Gate_Ruf_2024}, of which Josephson junction (JJ)-based superconducting circuits \cite{Thin-film_Lowell_2016, Quantum-classical_McDermott_2018, Microwave_Graninger_2023, Switch_Pechal_2016, On-chip_Naaman_2016, Josephson_Naaman_2017, Fast_Chang_2020, Modular_Wu_2024, General_Chapman_2016} have the unique advantage of easy integration with other JJ-based circuit elements. These circuits are often actuated by threading a dynamic magnetic flux into loops containing JJs, while simultaneously being biased by a static flux. However, because it's hard to engineer a unity inductive coupling factor between the flux control lines and the flux sensitive loops under tight space constraints, the currents injected into the cryostat need to be proportionally larger \cite{Fast_Gargiulo_2021, High-fidelity_Lu_2023}. In addition to dissipating more power, the large applied currents also exacerbate potential crosstalk between nearby devices, limiting their density on a single chip. Additionally, sensitivity to the uncontrollable background fluxes leads to variable and unreliable performance across each thermal cycle.

Here, we demonstrate a switch that is well adapted for densely integrated circuits embedded in superconducting environments. Our switch implements an efficient control scheme that consists of a persistent "set-and-forget" current bias and direct current (DC) actuation.
We first show that the persistent current, corresponding to up to hundreds of flux quanta, can be reliably and precisely trapped within $\text{200 μs}$ and decays by less than $1\%$ per day. We then show that, once biased, the switch can be actuated within a nanosecond to support a variety of resonator readout and multiplexing schemes.
The switch's on/off contrast is $>\text{20 dB}$ for nearly $\text{2 GHz}$ bandwidth, comparable to commercial ferrite isolators. Finally, its power handling is compatible with typical qubit readout tones, amplifier pumps and other microwave signals \cite{Engineering_Krinner_2019}.

\begin{figure*}
    \centering
    \includegraphics[width = 180mm]{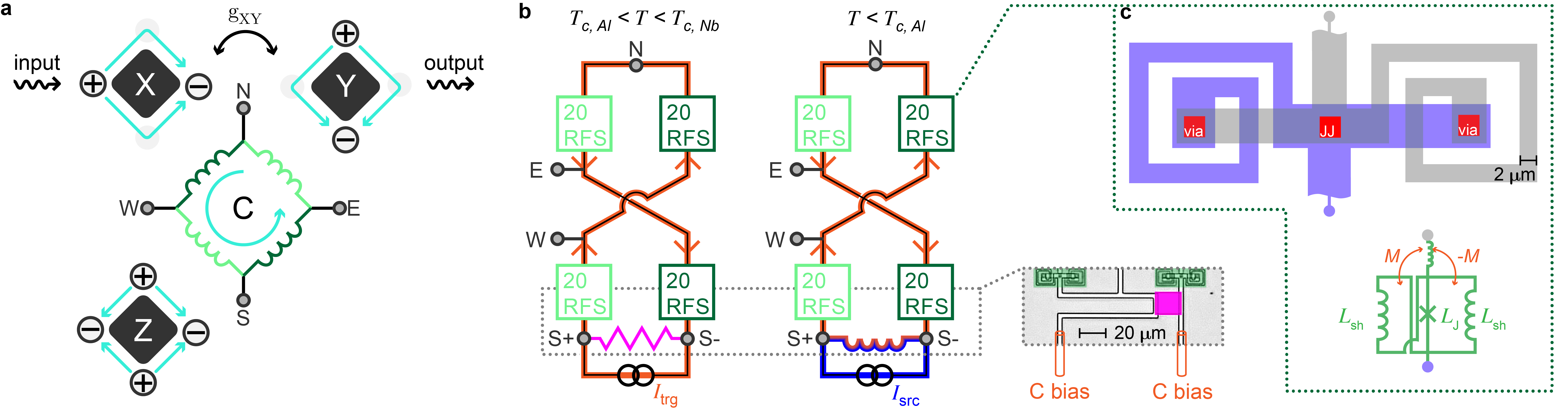}
    \caption{A current controlled switch concept and design.
    (a) The inductive Wheatstone bridge and its eigenmodes, $\text{X, Y, Z}$, expressed in terms of node flux variables $\phi_{\text{X}}= (\phi_{\text{W}} - \phi_{\text{E}})/\sqrt{2}$, $\phi_{\text{Y}}=(\phi_{\text{N}}  - \phi_{\text{S}}) /\sqrt{2}$ and $\phi_{\text{Z}}=(\phi_{\text{N}}  +\phi_{\text{S}}  - \phi_{\text{E}}  - \phi_{\text{W}})/2$\, \cite{Flurin_2014}. The currents associated with these modes are represented with the cyan arrows. In addition, a circulating current $I_{\text{C}}$ (central cyan arrow) is associated with the accumulative phase $\phi_{\text{C}}$ around the bridge. Using both $I_{\text{Z}}$ and $I_{\text{C}}$ as control currents, the total currents in the inductors are the same for opposite inductors but different for adjacent ones, as indicated by their light and dark green colors. The differential $\text{X}$ and $\text{Y}$ modes are connected to the input and output ports respectively.
    (b) The circuit schematics for the bridge concept in (a) shown at two temperatures. The diamond-shaped bridge is twisted into a figure-8 shape to minimize the flux sensitivity to any uniform background magnetic field. The rf-SQUID arrays, labeled as "20 RFS", implement the tunable inductors of the bridge.
    A PCS directs which path the source current $I_{\text{src}}$ takes.
    When the device temperature is such that the aluminum patch (magenta) is resistive, the south node $\text{S}$ splits into $\text{S}+$ and $\text{S}-$ nodes, through which the intended-to-trap current $I_{\text{src}}=I_{\text{trg}}$ flows into the bridge.
    When superconducting, the aluminum patch acts as a low inductance shunt, approximately collapsing the $\text{S}+$ and $\text{S}-$ nodes to the $\text{S}$ node. In this state, the $I_{\text{C}}$ consists of the trapped current ($\approx I_{\text{trg}}$) and the contribution from $\phi_{\text{ext}}$.
    Inset: An aluminum patch (magenta) closes the superconducting loop of the bridge.
    An antisymmetric rf-SQUID (c) is one in a series of 20 SQUIDs that form a tunable inductor \cite{Miniature_Zimmerman_1971}. In this compact design, there are two wiring layers shown in purple and gray, and they are connected by three parallel paths: 2 vias and a JJ. The middle trace inside the rf-SQUID and the trace connecting to the neighboring rf-SQUIDs interact with the spiral inductors ($L_{\text{sh}}$) through mutual inductance $M$ and $-M$, canceling the effect of each other.}
    \label{fig:1}
\end{figure*}
\begin{figure*}[t!]
    \centering
    \includegraphics[width = 180mm]{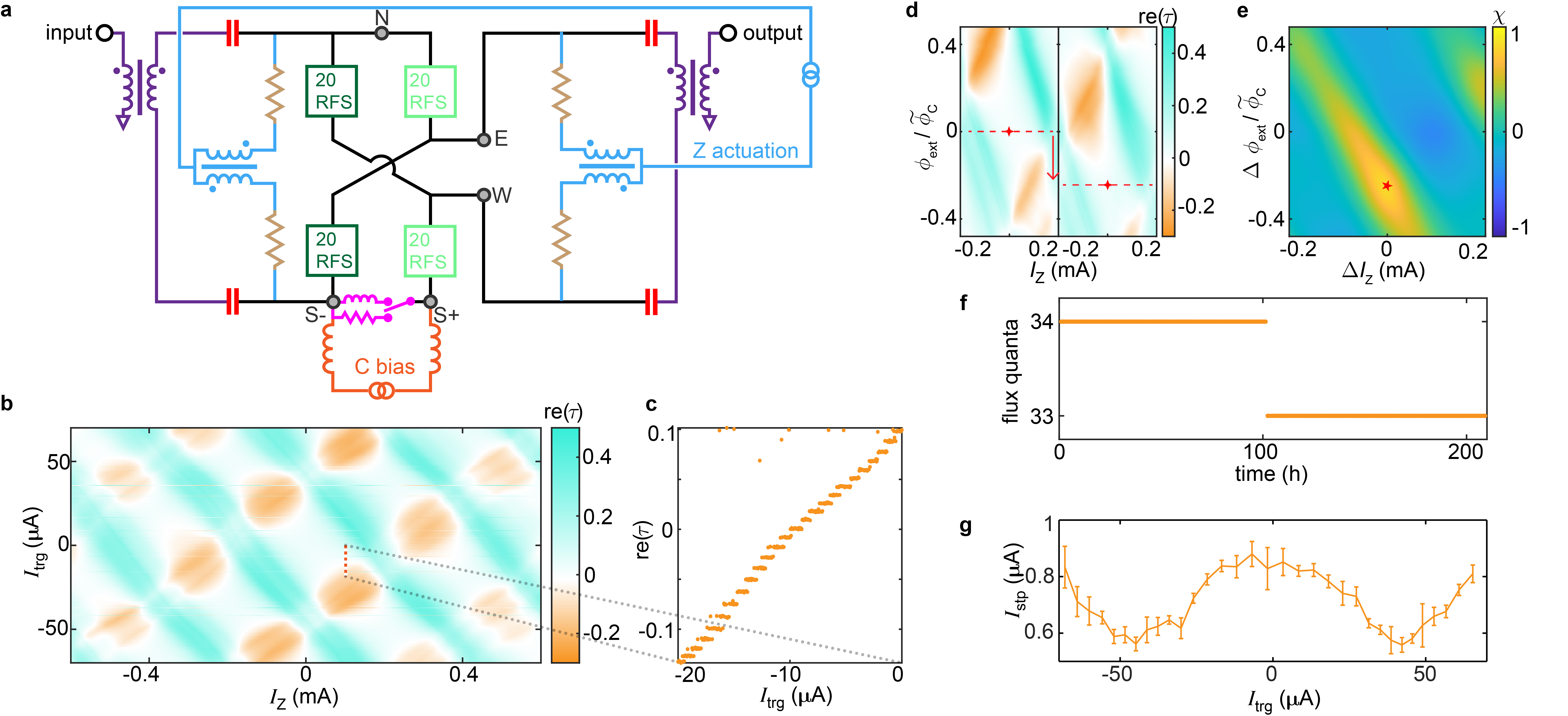}
    \caption{Demonstrating the control scheme with transmission measurements.
    (a) The switch's circuit diagram shows the bridge, the current control lines and the microwave ports for transmission measurements including the C bias (orange) and the Z actuation (blue). The bridge loop is closed with a heat activated PCS (magenta). The input and output ports are connected to the $\text{X}$ and $\text{Y}$ modes via baluns (purple) . The $10\ \text{m}\Omega$ Au resistors (brown) in the $\text{Z}$ lines and capacitors (red) break unwanted superconducting loops.
    (b) Transmission measurements sweeping $I_{\text{trg}}$ and the $I_{\text{Z}}$ are obtained by attempting to trap $I_{\text{trg}}$ using the PCS, followed by a sweep of $\tau(I_{\text{Z}})$. 
    (c) A vertical linecut of (b) at $I_{\text{Z}}=0.1\ \text{mA}$ reveals the persistent current trapped in the bridge forms quantized steps, corresponding to integer numbers of trapped flux.
    (d) Two measurements of $\tau(I_{\text{Z}},\phi_{\text{ext}})$ are shown at different values of trapped flux, $\phi_{\text{C}}=0$ (left) and $\phi_{\text{C}}=22$ (right), where $\phi_{\text{ext}}$ is normalized by its experimentally determined period
    $\tilde\phi_{\text{C}}$. The red crosses with dashed lines label the centers of the grid structures (points about which $\tau$ is most approximately an even function of both arguments) showing a shift of the grid pattern in the $\phi_{\text{ext}}$ axis. 
    (e) The figure shows the normalized correlation $\chi$ between the two sweeps of $\tau(I_{\text{Z}},\phi_{\text{ext}})$ in (d), where its maximum (red star) is used to extract the shift in $\phi_{\text{ext}}$ and the corresponding change in $I_{\text{C}}$.
    (f) This figure shows the result of repeatedly extracting the shift in $\phi_{\text{ext}}$ once an hour after intentionally trapping 34 flux quanta. The number of trapped flux remained constant until a jump (of unknown cause) occurred at the 104th hour.
    (g) The width of the steps shown in Fig.~\ref{fig:2}b represents the amount of $\text{C}$ bias required to increment the trapped flux.
    The step width $I_{\text{stp}}$ as a function of $I_{\text{trg}}$ changes periodically, illustrating the bridge's current dependent differential inductance.
    }
    \label{fig:2}
\end{figure*}

\section*{An inductive Wheatstone bridge controlled by currents}\label{sec2}
The efficient control scheme and the desirable performance of the switch are enabled by an inductive Wheatstone bridge composed of four current-tunable inductors that are in turn composed of series arrays of inductively shunted Josephson junctions, also known as radio-frequency superconducting quantum interference devices (rf-SQUIDs), with Josephson inductance $L_{\text{J}}$ and shunt inductance $L_{\text{sh}}$. In this circuit, each of the control and signal currents has a dedicated current pattern (Fig.~\ref{fig:1}a). This mapping can be understood from the Hamiltonian of the bridge $H$, written in flux variables $\phi$ normalized to flux quanta ($\phi_0=\hbar/2e$), which provides 3 eigenmodes $\text{X, Y, Z}$ and a constraint $\text{C}$. The associated current patterns $I_k=(1/\phi_0)\ \partial H / \partial \phi_k, \ k\in\{\text{X,Y,Z,C}\}$ are orthogonal and mapped to the input signal, output signal, actuation and bias respectively \cite{Analog_Bergeal_2010, On-Chip_Kerckhoff_2015}. The transmission of the bridge is then determined by the coupling rate $g_{\text{XY}}$ between the $\text{X}$ and $\text{Y}$ modes, which is controlled by $I_{\text{Z}}$ and $I_{\text{C}}$. This dependence can be expressed by the lowest-order (in $\phi_{\text{X}}$ and $\phi_{\text{Y}}$) interaction term in the Hamiltonian, where $ H_{\text{int}}/\hbar= g_{\text{XY}}  \phi_{\text{X}} \phi_{\text{Y}}$,
\begin{align}
   g_{\text{XY}} &\propto \sin(\phi_{\text{Z}}/N) \sin(\phi_\text{C}/4N), 
    \label{eqn:1}
\end{align}
and $N = 20$ is the number of rf-SQUIDs in an array (see Methods).
Similarly, the lowest order unwanted Kerr nonlinearity can be expressed as $ H_{\text{Kerr}}/\hbar = K_{\text{XY}}O(\phi_{\text{X}}^2 \phi_{\text{Y}}^2 ,\phi^4_{\text{X}},\phi^4_{\text{Y}})$, where
\begin{align}
   K_{\text{XY}}\propto \cos(\phi_{\text{Z}}/N) \cos(\phi_\text{C}/4N).
    \label{eqn:2}
\end{align}
To maximize the tunability of the desirable linear interaction (Eq.~\ref{eqn:1}) and minimize the unwanted Kerr nonlinearity (Eq.~\ref{eqn:2}), $\phi_\text{C}/2\pi$ should be set to $N (\pm 1 + 4j), j\in \mathbb{Z}$. 
Because $\phi_\text{C}=2\pi j + \phi_{\text{ext}}$, there is a unique opportunity to set $\phi_{\text{ext}} = 0$ and trap $I_{\text{C}}$ such that $j=\pm N$. We note that in the presence of stray inductance $L_{\text{str}}$ in the bridge, the optimal $j\rightarrow\pm N+L_{\text{str}}/2L_{\text{sh}}$, but the optimal $I_{\text{C}}$ remains unchanged (see Methods).

This ``set and forget'' $I_{\text{C}}$ is trapped in a fully superconducting path of the bridge, which is formed by the niobium linear inductors inside the rf-SQUIDs, the bridge wiring, and an aluminum patch with a lower critical temperature $T_{\text{C}}$ (Fig.~\ref{fig:1}b). 
This aluminum patch acts as a heat-activated persistent current switch (PCS), which can momentarily turn normal to change the number of flux quanta trapped in the bridge, without driving the higher $T_{\text{C}}$ niobium wiring normal\cite{Characterization_Leuthold_1994, Thin-film_Balchandani_2005}. 
To control the PCS, we use the $\text{Z}$ actuation lines that are galvanically connected to the bridge as the local heating circuit, because they can reliably turn normal when applying a current above $4.8\ \text{mA}$. Our protocol quickly ramps up $I_\text{Z}$ above this threshold and down to $\text{0}$ in $\text{200 μs}$, while applying the target $\text{C}$ bias ($I_{\text{trg}}$). This process generates heat that propagates to the aluminum patch, turning it normal as well. In this state, the previously trapped $I_{\text{C}}$ is dissipated, and $I_{\text{trg}}$ flows directly into the bridge. When the temperature of the aluminum patch cools down to below $T_{\text{C, Al}}$, a discrete value of $I_{\text{C}}\approx I_{\text{trg}}$ (associated with an integer number of flux quanta) is trapped in the bridge. We note that after this protocol, the shunt inductance of the PCS $L_{\text{PCS}}$ is much smaller than the bridge inductance. Thus, when the PCS is superconducting, the C bias current $I_{\text{src}}$ acts as a flux bias with $\phi_{\text{ext}} = L_{\text{PCS}} I_{\text{src}}/\phi_0$.

In addition to providing non-linearity and completing the superconducting loop, the rf-SQUIDs are also designed for microwave performance.
They have a small footprint where a JJ is connected to two linear inductors folded in spirals across two wiring layers (Fig.~\ref{fig:1}c). The spirals are counter-wound on the opposite sides of the junction, such that the rf-SQUID does not couple to the flux gradient created by the current flowing through the junction itself. The inductance ratio between the spiral inductors and the junction is $\beta = L_{\text{sh}}/L_{\text{J}} =L_{\text{sh}}I_0/2\pi\phi_0=  1.2$, where $I_0$ is the critical current of the junction, and is well below the hysteresis threshold of $2$. This condition guarantees a unique flux distribution across the rf-SQUIDs at any given $I_{\text{C}}$, allowing the PCS to control the $\text{C}$ bias without affecting any individual rf-SQUID. Lastly, the rf-SQUIDs are connected in arrays of $N=20$ on each arm of the bridge to provide a better combination of power handling and impedance matching to $ \text{50 } \Omega $ ports.

\section*{Transmission measurements demonstrating reliable and precise switch controls}\label{sec3}
To demonstrate the control scheme and characterize the switch's performance, we connect the four nodes of the bridge circuit (Fig~\ref{fig:1}a) to ports associated with the $\text{X/Y/Z/C}$ current patterns (Fig.~\ref{fig:2}a). 
The $\text{C}$ bias current is injected into and extracted from the bridge through the $\text{S+}$ and $\text{S-}$ nodes, and the $\text{Z}$ actuation current is injected into $\text{N}$ and $\text{S-}$ nodes and extracted from $\text{E}$ and $\text{W}$ nodes.
In both control lines, spiral inductors are used as low-pass filters to block microwave signals from leaving via the ports. In the Z actuation lines, they are also counter-wound to achieve a symmetry-dependent filtering: the differential signal currents ($\text{X}$ and $\text{Y}$) are filtered more effectively by the mutual inductance, while the common-mode control current ($\text{Z}$) is filtered less.
The differential $\text{X}$($\text{Y}$) mode on the left(right) side of the figure-8 shaped bridge is connected to the single-ended input(output) port via a balun. 
Finally, the resistors and capacitors are added to break unwanted superconducting loops and thus prevent them from trapping flux.

In Fig.~\ref{fig:2}b, we measure the microwave transmission $\tau(I_{\text{Z}}, I_{\text{trg}})$, and observe several features expected from the bridge model (Eq.~\ref{eqn:1}), along with some deviation from this simple model. Most notably, $\tau(I_{\text{Z}}, I_{\text{trg}})$ is periodic in both control parameters, as it oscillates between positive and negative values, mirroring the periodicity in $g_{\text{XY}}(I_{\text{Z}}, I_{\text{trg}})$ because $\tau$ is a monotonic function of $g_{\text{XY}}$.  
Moreover, we observe that $\tau(I_{\text{trg}})$ exhibits discrete steps, whose width $I_{\text{stp}}$ represents the amount of $I_{\text{trg}}$ required to increment the discrete trapped flux (Fig.~\ref{fig:2}c). Features that are not explained by Eq.~\ref{eqn:1}  include the skew in the periodic grid, which indicates a coupling between $I_{\text{Z}}$ and $\phi_{\text{C}}$, and a splitting of the transmission maxima and minima, which we believe arises from sensitivity of the antisymmetric rf-SQUIDs to a uniform background magnetic field. We believe that both of these effects can be reduced in a future device with straight-forward changes to the circuit layout. Finally, we note that the device has about 6~dB of insertion loss from the poor input match to the 50 $\Omega$ ports. This loss arises because we designed the devices for direct coupling to cavities (similar to Ref.~\onlinecite{Integrating_Zhao_2023}) rather than to these ports. We have achieved insertion loss below 1~dB in similar devices, when we optimized for coupling to the 50 $\Omega$ ports \cite{rosenthal_efficient_2021}. 

Having observed the periodicity of $\tau$, we now use it to study the stability of the persistent current. We first demonstrate the ability to measure the persistent current without changing it.  We intentionally trap two different numbers of flux quanta, and make use of the ability to tune $\phi_{\text{ext}}$ without changing $I_{\text{C}}$ to measure $\tau(I_{\text{Z}},\phi_{\text{ext}})$ (Fig.~\ref{fig:2}d). The two sweeps are similar except for a shift in $\phi_{\text{ext}}$, which can be interpreted as the external flux required to cancel the effect of the different numbers of trapped flux quanta. We extract this shift $\Delta \phi_{\text{ext}}$ from the maximum of the normalized correlation $\chi(\Delta I_{\text{Z}}, \Delta \phi_{\text{ext}})$ between the grids (Fig.~\ref{fig:2}e). With this ability to measure the persistent current, we then monitor it for a long period of time to detect jumps. We trap the current for optimal switch operation and then repeatedly extract this $\Delta \phi_{\text{ext}}$. In an observation run of a week with hourly monitoring, only a single flux jump occurred at about $\text{100 h}$ (Fig.~\ref{fig:2}f). 
The low jump rate suggests that the C bias remains stable for significantly longer than the transmons' hour-scale parameter-stable interval \cite{Klimov_Fluctuations_2018} or the daily calibration schedule in large-scale quantum processors \cite{IBM_CalibrationJobs2025, vtt_qx_calibration_2025}.

We further use the discretized nature of trapped flux to directly measure the differential inductance of the bridge loop and its stray inductance $L_{\text{str}}$, in contrast to inferring them from a detailed microwave model.
These measurements allow us to characterize circuit parameters more robustly by simply counting steps of $\tau$ (visible in Fig.~\ref{fig:2}c).
Both the number of steps in $\tau(I_{\text{C}})$ and the width of each step $I_{\text{stp}}$ are obtained by dividing $\tau(I_{\text{trg}})$ into groups that correspond to the same trapped flux (see Methods).
By extracting $I_{\text{stp}}$, we measure the differential inductance ($\phi_0 / I_{\text{stp}}$) of the bridge loop and observe that it varies by a factor of $1.5$ within a period (Fig.~\ref{fig:2}g). Because this factor relates to $\max(|\tau|)$ monotonically, it can be used to screen devices or compare different iterations of the same design. 
We note that this measurement relies only on a precisely controlled DC current $ I_{\text{stp}}$ and a constant of nature $\phi_0$.
In addition, we determine that $120$ flux quanta are required to modulate the transmission by one period, indicating the presence of linear stray inductance $L_{\text{str}}=(120-4N)L_{\text{sh}}=1.3\ \text{nH}$ in series with the bridge (see Methods). Again, this result only relies on the simulated value of $L_{\text{sh}}$ and an integer determined experimentally by counting discrete well-resolved steps.

\section*{Broadband and versatile microwave performance}\label{sec4}

\begin{figure}[h]
    \centering
    \includegraphics[width= 86 mm]{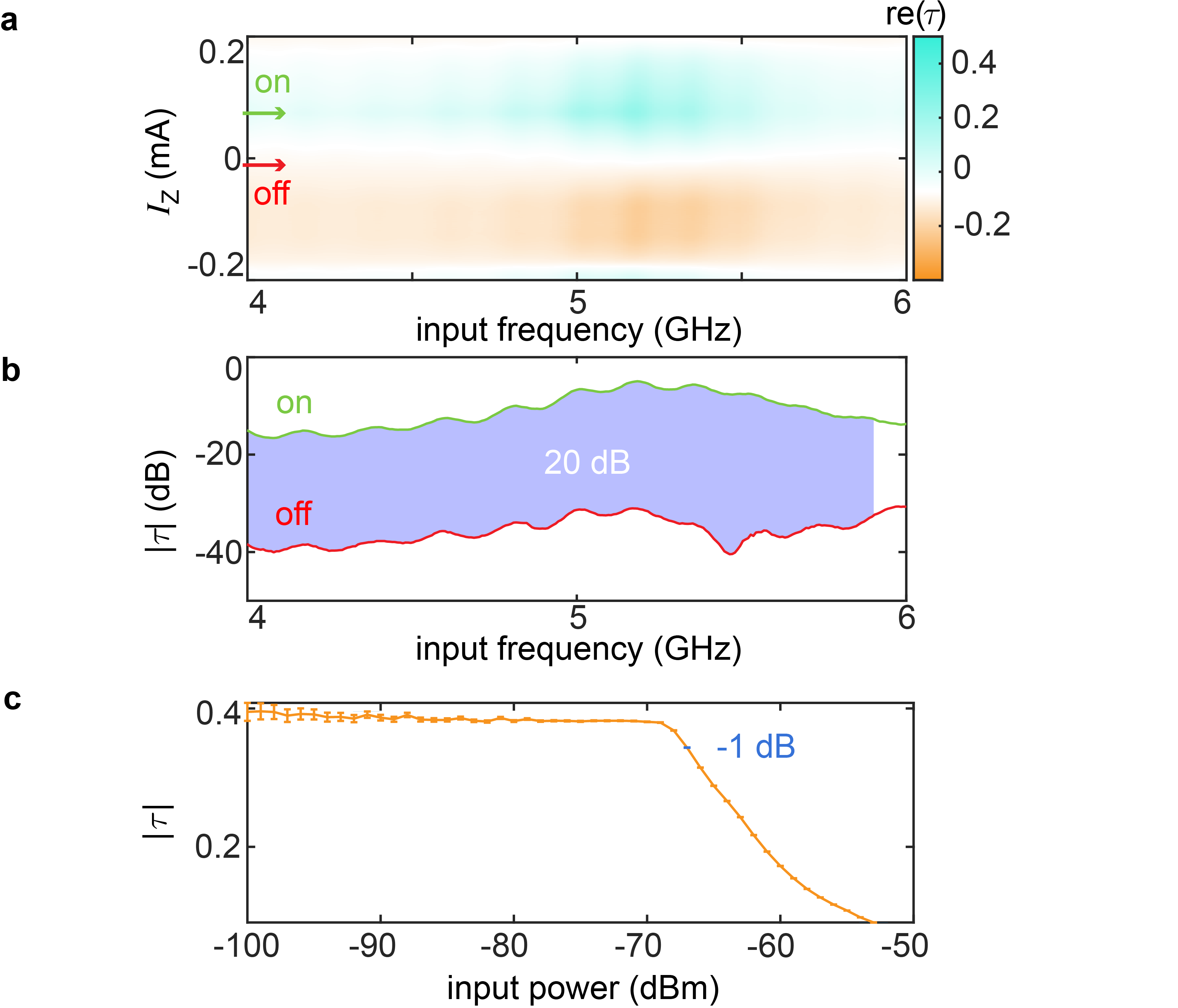}
    \caption{Microwave performance of the switch.
    (a) Transmission as a function of input frequency and $I_{\text{Z}}$, with $I_{\text{trg}} =  2\text{0 μA}$, shows regions of $\text{re}(\tau)>0$ (cyan), $\text{re}(\tau)\approx0$ (white) and $\text{re}(\tau)<0$ (orange). The red and green arrows label the $I_{\text{Z}}$ corresponding to the on and off modes.
    Figure (b) shows $|\tau|$ as a function of input frequency for the on and off modes. The purple shading indicates the region for which the difference between on and off transmission exceeds  $\text{20 dB}$.
    (c) The figure shows $|\tau|$ as function of input power of a $\text{5.1 GHz}$ tone for the on-mode bias point with the $\text{1 dB}$ compression point (annotated in blue) at $\text{-67 dBm}$.}
    \label{fig:3}
\end{figure}
 
Because our switch uses circuit symmetry rather than frequency mismatch to achieve isolation, it operates over a broad range of input frequencies. In addition, it remains linear for signal powers in excess of $\text{100 pW}$, much larger than typical superconducting qubit readout powers.
We determine the on/off ratio and its bandwidth from frequency-dependent $\tau$ measurements, where $I_{\text{C}}$ is fixed and $I_{\text{Z}}$ is swept to tune $\tau$ from its maximum to minimum (Fig.~\ref{fig:3}a).
We select two values of  $I_{\text{Z}}$ as the optimal on and off states, achieving $>20\ \text{dB}$ contrast over a $1.9\ \text{GHz}$ signal bandwidth (Fig.~\ref{fig:3}b). 
We then measure the power handling of the switch by sweeping the signal input power and extracting the power at which the $|\tau|$ drops by $\text{1 dB}$ (Fig.~\ref{fig:3})c. The $\text{1-dB}$ compression point occurs at 200~pW at the device, whereas typical qubit readout powers are about 1 million times smaller.

\begin{figure}
    \centering
    \includegraphics[width= 86 mm]{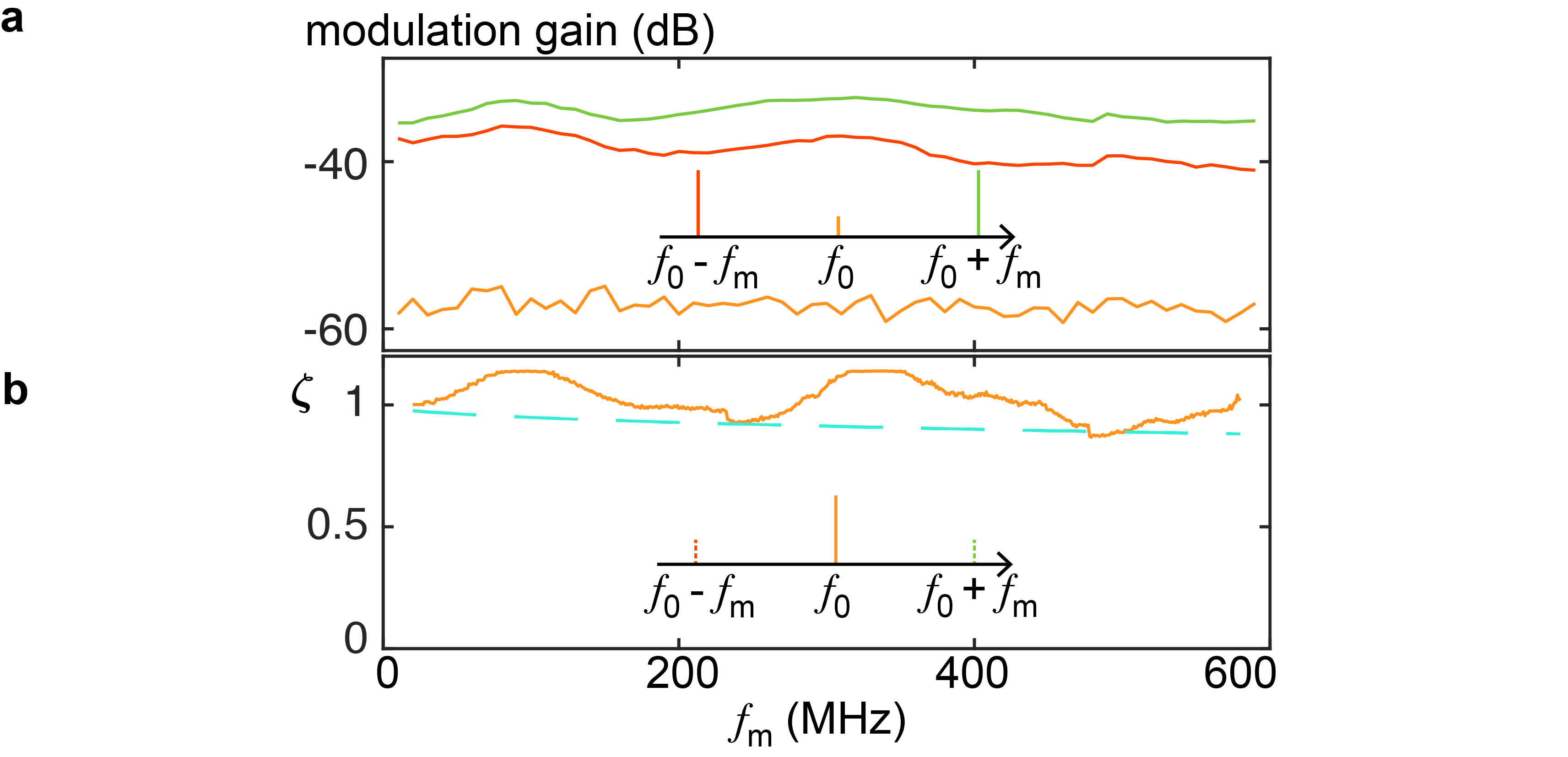}
    \caption{Sideband generation and modulation bandwidth. 
    (a) The modulation gain of the first-order modulation sidebands (red/green) and the $\text{5.1 GHz}$ carrier tone (orange) is plotted versus modulation frequency. The $|\tau|$ in all other sidebands is less than carrier feedthrough. The inset conceptually illustrates the color scheme in the main figure.
    (b) Extracted scaling factor $\zeta$ as a function of modulation frequency $f_{\text{m}}$ is shown as the orange line. Additionally, the cyan line shows the expected frequency dependent attenuation of the cryostat cable ($\text{1 dB}$ at $\text{5 GHz}$).
    The inset is similar to that of (a), except that the $|\tau|$ of the feedthrough is the dominant peak with $I_{\text{trg}}=0$. The first order sidebands are further filtered by a low-pass filter, and are represented by the dashed lines.
    }
    \label{fig:4}
\end{figure}

Finally, we operate our switch as a three-wave-mixer and show that its modulation bandwidth is larger than $600\ \text{MHz}$. 
Because this bandwidth is substantially larger than that of the electronics used to measure the transmitted signal, we bound the modulation bandwidth by tracking the $\tau$ of modulation sidebands and of an unmodulated carrier.
First, we select an $I_{\text{trg}}$ that minimizes the transmission of a 5.1~GHz carrier tone and apply a small oscillatory actuation current with amplitude $I_{\text{Z,0}} = 0.05 \ \tilde I_{\text{Z}}$, creating the modulation sidebands on the 5.1~GHz carrier. In Fig.~\ref{fig:4}a, we plot the modulation gain of the carrier and the modulation sidebands versus modulation frequency $f_\mathrm{m}$. There is not an apparent 3~dB bandwidth visible in the average amplitude in the two first order sidebands, however the gain in the upper and lower sidebands differs by an amount that grows with $f_\text{m}$. We attribute this discrepancy to the frequency-dependent $\tau$ of the switch itself, which is particularly important when the sidebands are detuned from each other by more than a gigahertz, comparable to the switch's signal bandwidth. 
We can distinguish this effect from the combination of the intrinsic modulation bandwidth and the frequency-dependent attenuation of the Z actuation lines by measuring transmission (at the fixed carrier frequency) versus $I_{\text{Z,0}}$ as it becomes comparable to $\tilde{I}_Z$.
We anticipate that the $\tau$ of the carrier will follow the expression:
\begin{equation} \label{eqn:bw}
    \text{re}(\tau) = \frac{c_0}{2} + \sum_n c_n J_0 \left[n\zeta(f_m) I_{\text{Z,0}}/\tilde{I}_Z\right],
\end{equation}
where $J_0$ is the Bessel function of the first kind, $\{c_n\}$ is the set of coefficients that can be independently obtained from the cosine expansion of transmission with static $I_{\text{Z}}$, and $\zeta$ is the scale introduced to account for the intrinsic modulation bandwidth and the frequency-dependent attenuation of the Z actuation lines.
We choose an operating point such that the switch is statically imbalanced and transmission is not minimized (Fig.~\ref{fig:4}b, inset), measure $\tau(f_{\text{m}}, I_{\text{Z,0}})$ and fit to Eq.~\ref{eqn:bw} with $\zeta(f_{\text{m}})$ as the only adjustable parameter.
We observe that $\zeta(f_{\text{m}})$ has not declined by $3\ \text{dB}$ as the modulation frequency reaches $600\ \text{MHz}$, indicating the modulation bandwidth is even wider. Furthermore, the slowly decaying $\zeta$ is approximately the attenuation expected from the coaxial cables in the cryostat. 

\section*{Conclusions}
We have described a superconducting switch controlled by a persistent current bias and a current actuation to achieve a lower power consumption and crosstalk compared to commonly used flux controls. The long-lived persistent current bias allows the switch to operate in a set-and-forget mode, reducing the need of the on-the-fly tuning. To this end, we show that our control scheme can precisely and reliably trap and re-trap the desired amount of bias current, along with an efficient detection scheme for the trapped current. Our switch also has desirable performance, including $\text{20 dB}$ of on/off contrast over a $\text{1.9 GHz}$ bandwidth, 200 pW power handling and modulation bandwidth wider than $600\ \text{MHz}$. These qualities make our switch well-suited to a large-scale superconducting circuit. 

\section*{Methods}
\subsection{Bridge Hamiltonian}
In this section, we formulate the Hamiltonian of the bridge circuit. 
Each arm of the bridge consists of a series array of $N=20$ rf-SQUIDs, and thus there are 20 independent flux coordinates in one arm. But in the low-frequency limit (below the lowest internal resonance frequency of the arm), the state of the arm can be described by a single branch flux across the arm.
Assuming that all the rf-SQUIDs are identical and non-hysteretic, this flux drop is equally distributed among all rf-SQUIDs within the arm. 
Connecting the four arms to form the bridge, we can express $H$ in the branch flux variables \cite{Flurin_2014}:
\begin{equation}
    H = N \sum_{l\in\{\text{NW,SW,SE,NE}\}} -E_J  \cos(\frac{\phi_l}{N}) + \frac{\phi_0^2}{L_{\text{sh}}} ( \frac{\phi_l}{N})^2,
\end{equation}
where
\begin{equation}
    \begin{aligned}
        \phi_{\text{NE}} & = \phi_{\text{E}} - \phi_{\text{N}} + \frac{\phi_{\text{C}}}{4} \\
        \phi_{\text{SE}} & = \phi_{\text{S}} - \phi_{\text{E}} + \frac{\phi_{\text{C}}}{4} \\
        \phi_{\text{SW}} & = \phi_{\text{W}} - \phi_{\text{S}} + \frac{\phi_{\text{C}}}{4} \\
        \phi_{\text{NW}} & = \phi_{\text{N}} - \phi_{\text{W}} + \frac{\phi_{\text{C}}}{4}.
    \end{aligned}
\end{equation}
We can express $H$ in the $\{\text{X,Y,Z,C}\}$ basis (defined in the Fig.~\ref{fig:1} caption) by transforming into the eigenmode basis, defined as:
\begin{equation}
\begin{aligned}
    \phi_\text{X} &= \frac{1}{\sqrt{2}}(\phi_{\text{W}} - \phi_{\text{E}}) \\
    \phi_\text{Y} &= \frac{1}{\sqrt{2}}(\phi_{\text{N}}  - \phi_{\text{S}})  \\
    \phi_\text{Z} &= \frac{1}{2}(\phi_{\text{N}}  +\phi_{\text{S}}  - \phi_{\text{E}}  - \phi_{\text{W}}),
\end{aligned}
\end{equation}
and resulting in: 
\begin{equation}
\begin{aligned}
    H &= \frac{ \phi_0^2}{N L_{\text{sh}}} (\frac{\phi_{\text{C}}^2}{4} + 2\phi_\text{X}^2 + 2\phi_\text{Y}^2 + 4\phi_\text{Z}^2)\\
    &+  4N E_J [ \text{cos}(\frac{\phi_{\text{C}}}{4N}) \text{cos}(\frac{\phi_\text{X}}{\sqrt{2}N})\text{cos}(\frac{\phi_\text{Y}}{\sqrt{2}N})\text{cos}(\frac{\phi_\text{Z}}{N})\\
   &+ \text{sin}(\frac{\phi_{\text{C}}}{4N}) \text{sin}(\frac{\phi_\text{X}}{\sqrt{2}N})\text{sin}(\frac{\phi_\text{Y}}{\sqrt{2}N})\text{sin}(\frac{\phi_\text{Z}}{N})].
\end{aligned}
\end{equation}
Expanding this Hamiltonian to second order in $\phi_{\text{X}}$, and $\phi_{\text{Y}}$, we find the interaction term (Eq.~\ref{eqn:1}) and the Kerr term (Eq.~\ref{eqn:2}). Finally, we determine the current pattern associated with a particular mode or constraint $k$ on arm $l$ from the sign of $\partial \phi_l/ \partial \phi_k $, as seen in the expression for the mode current
\begin{equation}
    I_{k}=\frac{1}{\phi_0}\ \sum_{l}\frac{\partial H}{ \partial \phi_l} \frac{\partial \phi_l}{ \partial \phi_k}  , \ l\in\{\text{NW,SW,SE,NE}\}, k\in\{\text{X,Y,Z,C}\}.
\end{equation}

\subsection{Periods of $g_{\text{XY}}$ as function of $I_{\text{C}}$ and $\phi_{\text{C}}$}
We first find the period $\tilde I_{\text{C}}$ of $g_{\text{XY}}(I_{\text{C}})$ which satisfies $I_{\text{C}}(\phi_{\text{C}})+\tilde I_{\text{C}}=I_{\text{C}}(\phi_{\text{C}}+\tilde\phi_{\text{C}})$.
From the simple model presented in Eq.~\ref{eqn:1}, the period of the flux variable is $\tilde \phi_{\text{C}} = 8\pi N$, and $\tilde I_{\text{C}}$ can be calculated from
\begin{align*}
    \frac{I_{\text{C}}+ \tilde I_{\text{C}}}{I_0} 
    &=  \frac{1}{E_{\text{J}}}\frac{\partial H}{\partial \phi_{\text{C}}} \vert_{\phi_{\text{C}}\rightarrow \phi_{\text{C}}+\tilde\phi_{\text{C}}} \\
    &= \frac{\phi_{\text{C}}+\tilde\phi_{\text{C}}}{2 N\beta} \\
    &-  \cos(\frac{\phi_{\text{X}}}{\sqrt{2} N})\cos(\frac{\phi_{\text{Y}}}{\sqrt{2}N})\cos(\frac{\phi_{\text{Z}}}{N})\sin(\frac{\phi_{\text{C}}+\tilde \phi_{\text{C}}}{4N})\\
   &+ \sin(\frac{\phi_{\text{X}}}{\sqrt{2} N})\sin(\frac{\phi_{\text{Y}}}{\sqrt{2}N})\sin(\frac{\phi_{\text{Z}}}{N})\cos(\frac{\phi_{\text{C}}+\tilde \phi_{\text{C}}}{4N}) \\
   &= \frac{I_{\text{C}}}{I_{0}} + \frac{\tilde\phi_{\text{C}}}{2N\beta}.
\end{align*}
We find $\tilde I_{\text{C}}=\tilde\phi_{\text{C}} I_0/2N\beta =  4\pi I_0/\beta$, where $\beta = L_{\text{sh}}/L_{\text{J}}$.

In a more complete model that accounts for the stray geometrical inductance $L_{\text{str}}$ associated with the bridge wiring, $\tilde I_{\text{C}}$ remains unchanged while $\tilde \phi_{\text{C}}$ increases (\ref{fig:stray}). 
This effect can be observed from the inductance of each arm $L_{l}( l\in\{\text{NW,SW,SE,NE}\})$, because it has the same periods as $g_{\text{XY}}$. $L_{l}$ can be expressed as:
\begin{equation}
     L_{l}  = \phi_0\frac{\partial \phi_l}{\partial I_l} =  N \frac{\phi_0}{I_0}\frac{1}{\cos{\phi_{\text{J}}+\frac{2}{\beta}}} + \frac{L_{\text{str}}}{4},
\end{equation}
where $\phi_{\text{J}}$ is the phase drop across each JJ. Any increment of $I_{\text{C}}$ or $\phi_{\text{C}}$ that varies $\phi_{\text{J}}$ by $2\pi$ leaves $L_l$ unchanged, and thus defines the period of $I_{\text{C}}$ or $\phi_{\text{C}}$. Because the same current $I_{l}$ flows through both the rf-SQUIDs and the stray inductance in one arm, the arm inductance has the same period $\tilde I_{\text{C}}= 4\pi I_0/\beta$ as a single rf-SQUID. In contrast, $\tilde \phi_{\text{C}}$ is the sum of the phase drop across the rf-SQUIDs and the stray inductance, which increases to $\tilde\phi_{C}\rightarrow8\pi N + 4\pi L_{\text{str}}/L_{\text{sh}} $. To achieve the maximal $g_{\text{XY}}$, the bias condition remains $\phi_{\text{J}}=\pi/2$ for all rf-SQUIDs, resulting in $j\rightarrow\pm N+L_{\text{str}}/2L_{\text{sh}}$.

\subsection{Extracting number of steps and step sizes of $I_{\text{trg}}$}

To extract the number of steps and the size of each step $I_{\text{stp}}$, we compare $\tau(I_{\text{Z}})$ at different values of $I_{\text{trg}}$ (horizontal linecuts of Fig.~\ref{fig:2}b), over a range of $I_{\text{Z}}$. 
The similarity between two linecuts is determined by the normalized correlation $\chi_{mn}$ between them, labeled by the $m^{\text{th}}$ and $n^{\text{th}}$ index of $I_{\text{trg}}$. It is defined as $ \chi_{mn} = \sum_d \tau_{m}^*(d) \tau_{n}(d)/\sqrt{\sum_d|\tau_{m}(d)|^2} \sqrt{\sum_d|\tau_{n}(d)|^2}\leq 1 $, where $d$ indexes the values of $I_{\text{Z}}$ and $\chi_{mn}=1$ if the two linecuts are identical. We interpret $I_{\text{trg}}(m)$ and $I_{\text{trg}}(n)$ as having the same value of trapped flux if $\chi_{mn}$ is above an experimentally determined threshold.

To find this threshold, we first calculate $\chi$ between all pairs $(m,n)$. In \ref{fig:quantized}a, we plot a representative example where $\chi_{mn}$ clearly exhibits discretized steps in $n$. We then find the histogram of $\chi_{mn}$ and observe that the result shows well separated peaks (\ref{fig:quantized}b). The $0^{\text{th}}$ peak groups cases in which $I_{\text{trg}}(m)$ and $I_{\text{trg}}(n)$ are the most similar, which we interpret as having the same value of trapped flux. As shown in the figure, we determine the threshold between the $0^{\text{th}}$ and $1^{\text{st}}$ peaks to be $\chi = 0.9992$. From \ref{fig:quantized}a, we also note that the flux trapping has a stochastic component, particularly evident at the boundary between two steps. Consequently, we assign boundaries between steps (finding $I_{\text{stp}}$) such that we minimize the mean-square distance to the boundary of indices assigned to the $1^{\text{st}}$ histogram peak within a $0^{\text{th}}$ histogram step. Finally, by counting the indices that are obvious outliers (with $\chi< 0.9$) and therefore correspond to failures to trap the intended flux, we determine that our PCS protocol has a $2.3\%$ failure rate for a single trap procedure. But because we can subsequently measure the amount of trapped flux, the trap and measure procedure can be repeated until success, driving this failure rate down as low as desired.

\subsection{Device fabrication}
This device is manufactured with the NIST niobium trilayer process (\ref{fig:pic}) \cite{NIST_trilayer} with critical current density of the JJs of 1.6~μA/μm$^2$. Process steps were added to create both the aluminum layer for the PCS and the non-superconducting gold patches, where both of these metal layers are 200~nm thick.

\section*{Data availability}
The data that support the findings of this study are available from the corresponding author upon reasonable request.

\section*{Acknowledgments}
This work was supported by Q-SEnSE: Quantum Systems through Entangled Science and Engineering (NSF QLCI Award No. OMA-2016244). We thank D. Schmidt for helpful discussions. 

\bibliography{sn-bibliography}

\newpage
\setcounter{figure}{0}
\renewcommand{\figurename}{}
\renewcommand{\thefigure}{Extended Data Fig.~\arabic{figure}}
\begin{figure*}
    \centering
    \includegraphics[width = 85mm]{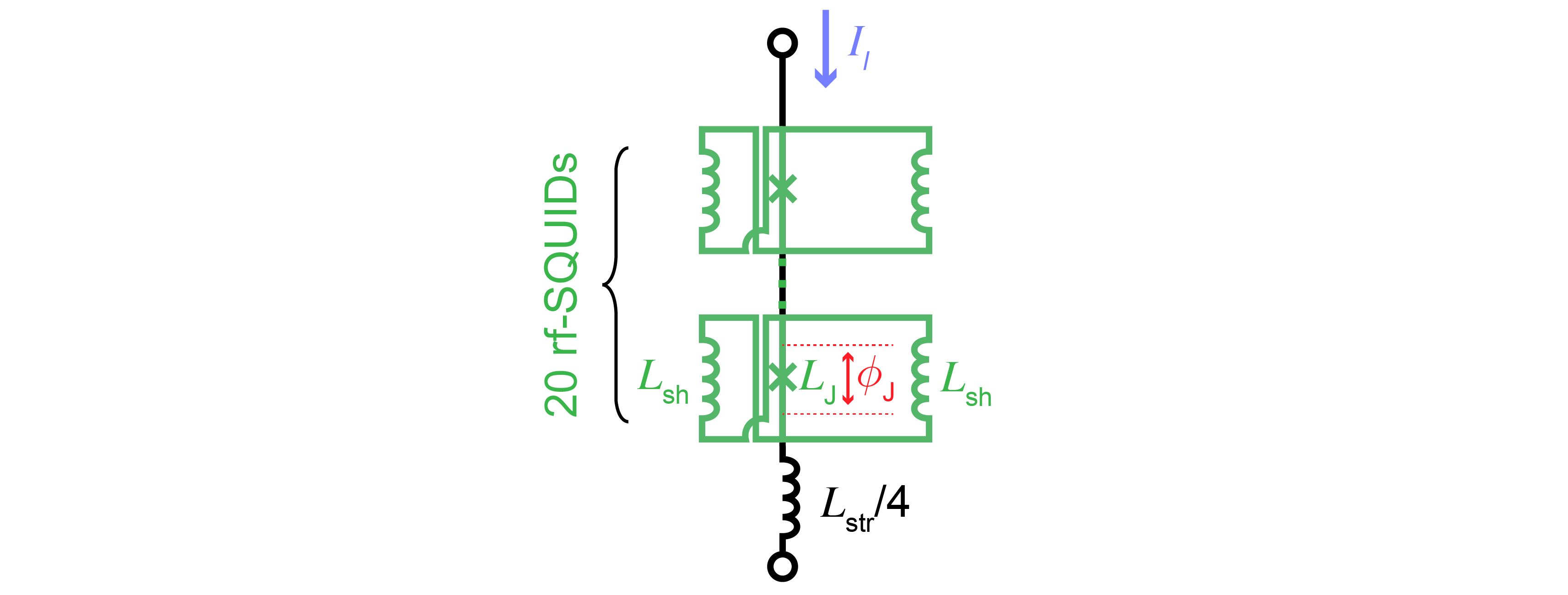}
    \caption{The circuit model of the bridge arm (labeled by $l$) includes 20 rf-SQUIDs and stray inductance $L_{\text{str}}/4$. The phase drop across each rf-SQUID is $\phi_{\text{J}}$, which is controlled by the current in the arm $I_{l}$.
    }
    \label{fig:stray}
\end{figure*}

\begin{figure*}
    \centering
    \includegraphics[width = 85mm]{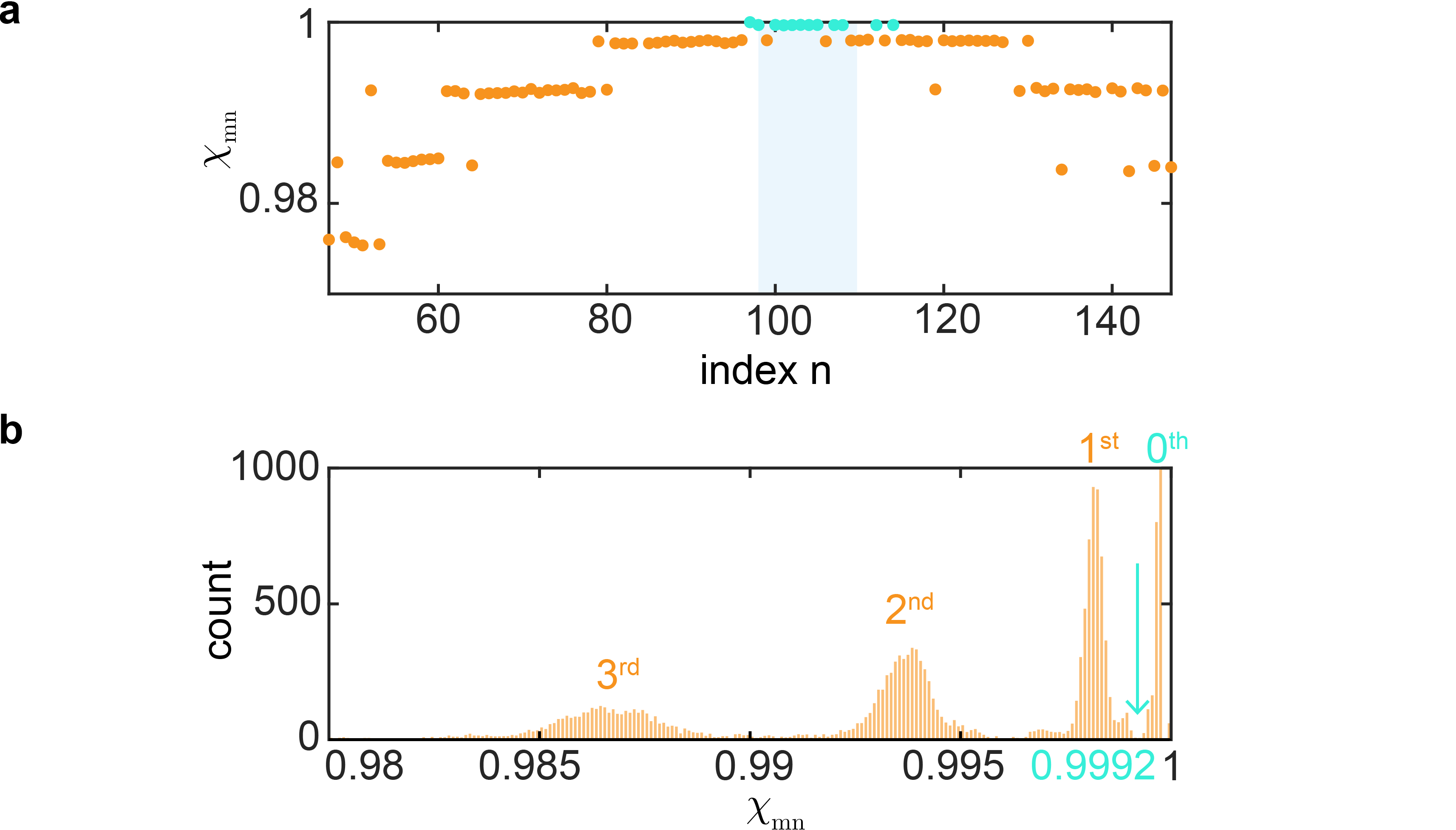}
    \caption{Grouping $I_{\text{trg}}$ by the corresponding trapped flux.
    (a) A representative example of $\chi_{mn}$ ($m=97$ and $47\leq n\leq147$) exhibits clear steps. The result of the grouping that contains $m=97$ is annotated in cyan. The corresponding $I_{\text{stp}}$ (defined in the main text) is shown as the blue region.
    (b) Histogram of $\chi_{mn}$ shows its first four peaks corresponding to $I_{\text{trg}}$ of the same flux, $\pm1$ flux, $\pm2$ flux and $\pm3$ flux. The threshold $\chi = 0.9992$ most effectively separates the $0^{\text{th}}$ and $1^{\text{st}}$ peaks.}
    \label{fig:quantized}
\end{figure*}

\begin{figure*}
    \centering
    \includegraphics[width = 85mm]{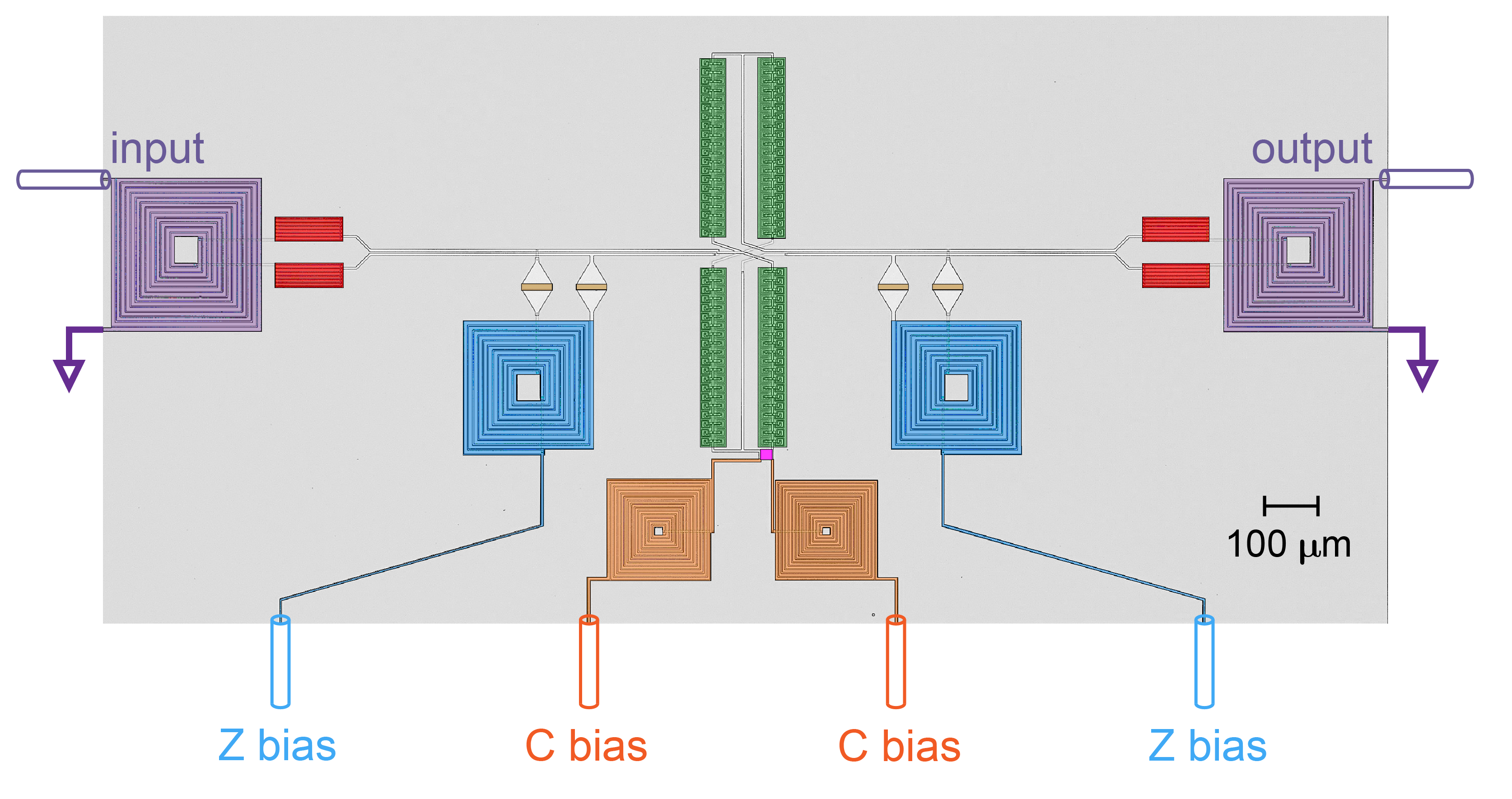}
    \caption{False color image of the switch. The color definition follows that of Fig.~\ref{fig:2}a.}
    \label{fig:pic}
\end{figure*}

\end{document}